\documentclass{aa}
\usepackage{graphicx}
\begin{document}

\title{The Insignificance of P-R Drag in Detectable Extrasolar
Planetesimal Belts}
\titlerunning{The Insignificance of P-R Drag}
\author{M. C. Wyatt
}
\offprints{M. C. Wyatt}
\institute{UK Astronomy Technology Centre, Royal Observatory,
              Edinburgh EH9 3HJ, UK\\
              \email{wyatt@roe.ac.uk}
             }
\date{Submitted 27 September 2004, Accepted 13 December 2004}

\abstract{This paper considers a simple model in which dust produced
in a planetesimal belt migrates in toward the star due to P-R drag
suffering destructive collisions with other dust grains on the way.
Assuming the dust is all of the same size, the resulting surface
density distribution can be derived analytically and depends only on
the parameter $\eta_0 = 5000 \tau_{eff}(r_0) \sqrt{M_\star/r_0}/\beta$;
this parameter can be determined observationally with the hypothesis
that $\beta=0.5$.
For massive belts in which $\eta_0 \gg 1$ dust is confined to the
planetesimal belt, while the surface density of more tenuous belts, in
which $\eta_0 \ll 1$, is constant with distance from the star.
The emission spectrum of dust from planetesimal belts at different
distances from different mass stars shows that the dust belts
which have been detected to date should have $\eta_0 \gg 1$;
dust belts with $\eta_0 \ll 1$ are hard to detect as they are
much fainter than the stellar photosphere.
This is confirmed for a sample of 37 debris disk candidates for
which $\eta_0$ was determined to be $>10$.
This means that these disks are so massive that mutual collisions
prevent dust from reaching the inner regions of these systems
and P-R drag can be ignored when studying their dynamics.
Models for the formation of structure in debris disks by the trapping
of particles into planetary resonances by P-R drag should be
reconsidered.
However, since collisions do not halt 100\% of the dust, this means
that in the absence of planetary companions debris disk systems should
be populated by small quantities of hot dust which may be detectable
in the mid-IR.
Even in disks with $\eta_0 \ll 1$ the temperature of dust emission is
shown to be a reliable tracer of the planetesimal distribution
meaning that inner holes in the dust distribution imply a lack of
colliding planetesimals in the inner regions.
}

\maketitle

\section{Introduction}
\label{s:intro}
Some 15\% of nearby stars exhibit more infrared emission than that
expected from the stellar photosphere alone
(e.g., Aumann et al. 1984).
This excess emission comes from dust in orbit around the stars
and its relatively cold temperature implies that it resides at
large distances from the stars, 30-200 AU, something which
has been confirmed for the disks which are near enough and
bright enough for their dust distribution to be imaged
(Holland et al. 1998; Greaves et al. 1998; Telesco et al. 2000;
Wyatt et al. 2004). 
Because the dust would spiral inwards due to Poynting-Robertson
(P-R) drag or be destroyed in mutual collisions on timescales
which are much shorter than the ages of these stars, the dust is
thought to be continually replenished (Backman \& Paresce 1993),
probably from collisions between km-sized planetesimals (Wyatt \& Dent
2002).
In this way the disks are believed to be the extrasolar
equivalents of the Kuiper Belt in the Solar System
(Wyatt et al. 2003).

These debris disks will play a pivotal role in increasing our
understanding of the outcome of planet formation.
Not only do these disks tell us about the distribution of
planetesimals resulting from planet formation processes,
but they may also provide indirect evidence of unseen planets in their
systems.
Models have been presented that show how planets can cause
holes at the centre of the disks and clumps in the azimuthal
distribution of dust, both of which are commonly observed features
of debris disks.
Many of these models require the dust to migrate inward
due to P-R drag to be valid;
e.g., in the model of Roques et al. (1994) the inner hole is caused by
a planet which prevents dust from reaching the inner system which would
otherwise be rapidly replenished by P-R drag (e.g., Strom, Edwards \&
Skrutskie 1993), and clumps arise in models when dust migrates inward
due to P-R drag and becomes trapped in a planet's resonances (Ozernoy
et al. 2000; Wilner et al. 2002; Quillen \& Thorndike 2002).
Alternative models exist for the formation of both inner holes
and clumps;
e.g., in some cases inner holes may be explained by the
sublimation of icy grains (Jura et al. 1998) or by the outward
migration of dust to the outer edge of a gas disk (Takeuchi \&
Artymowicz 2001), and clumps may arise from the destruction of
planetesimals which were trapped in resonance with a planet
when it migrated out from closer to the star (Wyatt 2003).

The focus of the models on P-R drag is perhaps not surprising, as
the dynamical evolution of dust in the solar system is undeniably
dominated by the influence of P-R drag, since this is the reason the
inner solar system is populated with dust (Dermott et al. 1994;
Liou \& Zook 1999; Moro-Mart\'{i}n \& Malhotra 2002).
However, there is no reason to expect that the physics dominating
the structure of extrasolar planetesimal disks should be the same as
that in the solar system.
In fact the question of whether any grains in a given disk suffer
significant P-R drag evolution is simply determined by how dense
that disk is (Wyatt et al. 1999).
It has been noted by several authors that the collisional
lifetime of dust grains in the well studied debris disks is shorter
than that of P-R drag (e.g., Backman \& Paresce 1993; Wilner et al.
2002; Dominik \& Decin 2003), a condition which means that P-R drag
can be ignored in these systems.

Clearly it is of vital importance to know which physical processes
are at play in debris disks to ascertain the true origin of these
structures.
In this paper I show that P-R drag is not an important physical
process in the disks which have been detected to date
because collisions occur on much shorter timescales meaning that
planetesimals are ground down into dust which is fine enough to be
removed by radiation pressure before P-R drag has had a chance to act.
In \S \ref{s:sm} a simple model is derived for the spatial distribution
of dust created in a planetesimal belt.
In \S \ref{s:em} this model is used to determine the emission spectrum
of these dust disks.
A discussion of the influence of P-R drag in detectable and detected
debris disks as well as of the implications for how structure in these
disks should be modelled and interpreted is given in \S \ref{s:disc}.

\section{Balance of Collisions and P-R Drag}
\label{s:sm}
In this simple model I consider a planetesimal belt at a distance
of $r_0$ from a star of mass $M_\star$ which is producing particles
all of the same size, $D$.
The orbits of those particles are affected by the interaction of the
dust grains with stellar radiation which causes a force which is
inversely proportional to the square of distance from the star,
and which is commonly defined by the
parameter $\beta=F_{rad}/F_{grav}$ (Burns et al. 1979; Gustafson 1994).
This parameter is a function of particle size and for large particles
$\beta \propto 1/D$.
The tangential component of this force is known as Poynting-Robertson
drag, or P-R drag.
This results in a loss of angular momentum from the particle's orbit
which makes it spiral in toward the star.
Assuming the particle's orbit was initially circular, the migration
rate is:
\begin{equation}
  \dot{r}_{pr} = -2\alpha/r, \label{eq:rpr}
\end{equation}
where $\alpha = 6.24 \times 10^{-4} (M_\star/M_\odot)\beta$
AU$^2$yr$^{-1}$ (Wyatt et al. 1999).

On their way in, dust grains may collide with other dust grains.
The mean time between such collisions depends on the dust density:
\begin{equation}
  t_{coll}(r) = t_{per}(r) / 4\pi \tau_{eff}(r), \label{eq:tcoll}
\end{equation}
where $t_{per} = \sqrt{(r/a_\oplus)^3(M_\odot/M_\star)}$ is the orbital
period at this distance from the star, and $\tau_{eff}$ is the effective
optical depth of the disk, or the surface density of cross-sectional
area of the dust (Wyatt et al. 1999).
If the collisions are assumed to be destructive then the distribution
of dust in the disk can be determined by considering the amount of
material entering and leaving an annulus at $r$ of width $dr$.
The steady state solution is that the amount entering the annulus due to
P-R drag is equal to that leaving due to P-R drag and that which is lost
by collisions (i.e., the continuity equation):
\begin{equation}
   d[n(r)\dot{r}_{pr}(r)]/dr = -N^{-}(r), \label{eq:cont}
\end{equation}
where $n(r)$ is the one dimensional number density (number of particles
per unit radius), and $N^{-}(r) = n(r)/t_{coll}(r)$ is the rate of
collisional loss of $n(r)$.
Since in a thin disk $\tau_{eff}(r) = 0.125 D^2 n(r)/r$, this
continuity equation can be solved analytically to find the
variation of effective optical depth with distance from the star
(Wyatt 1999):
\begin{eqnarray}
  \tau_{eff}(r) & = & \frac{\tau_{eff}(r_0)} 
    {1+4\eta_0(1-\sqrt{r/r_0})} \label{eq:prwcoll} \\
  \eta_0 & = &
    5000 \tau_{eff}(r_0)\sqrt{(r_0/a_\oplus)(M_\odot/M_\star)}/\beta
    \label{eq:eta0}
\end{eqnarray}
where this distribution has been scaled by the boundary condition
that at $r_0$, $\tau_{eff} = \tau_{eff}(r_0)$.

\begin{figure*}
  \centering
  \begin{tabular}{rlrl}
    \textbf{(a)} &
    \hspace{-0.6in} \includegraphics[width=3.2in]{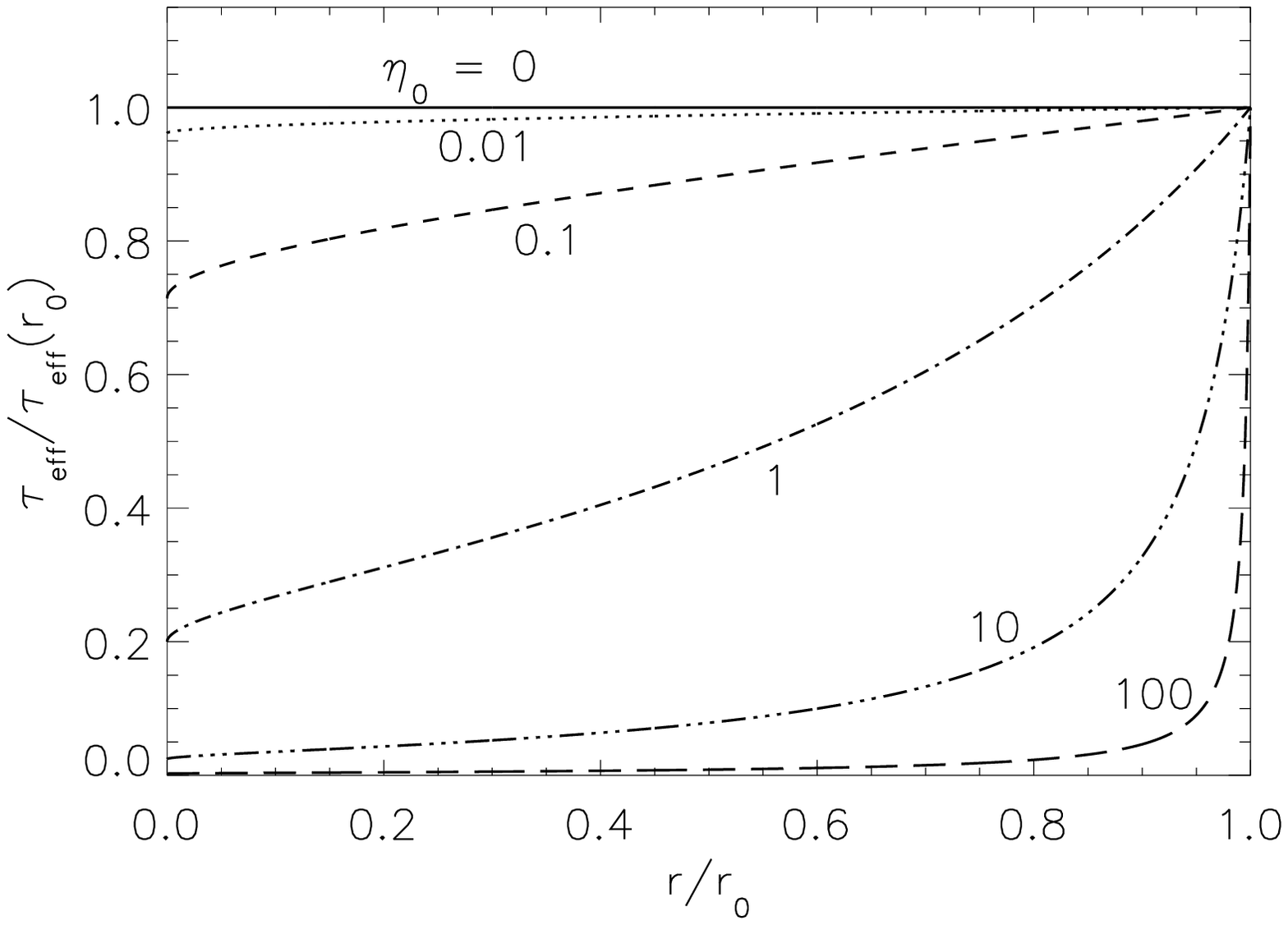} &
    \textbf{(b)} &
    \hspace{-0.3in} \includegraphics[width=3.2in]{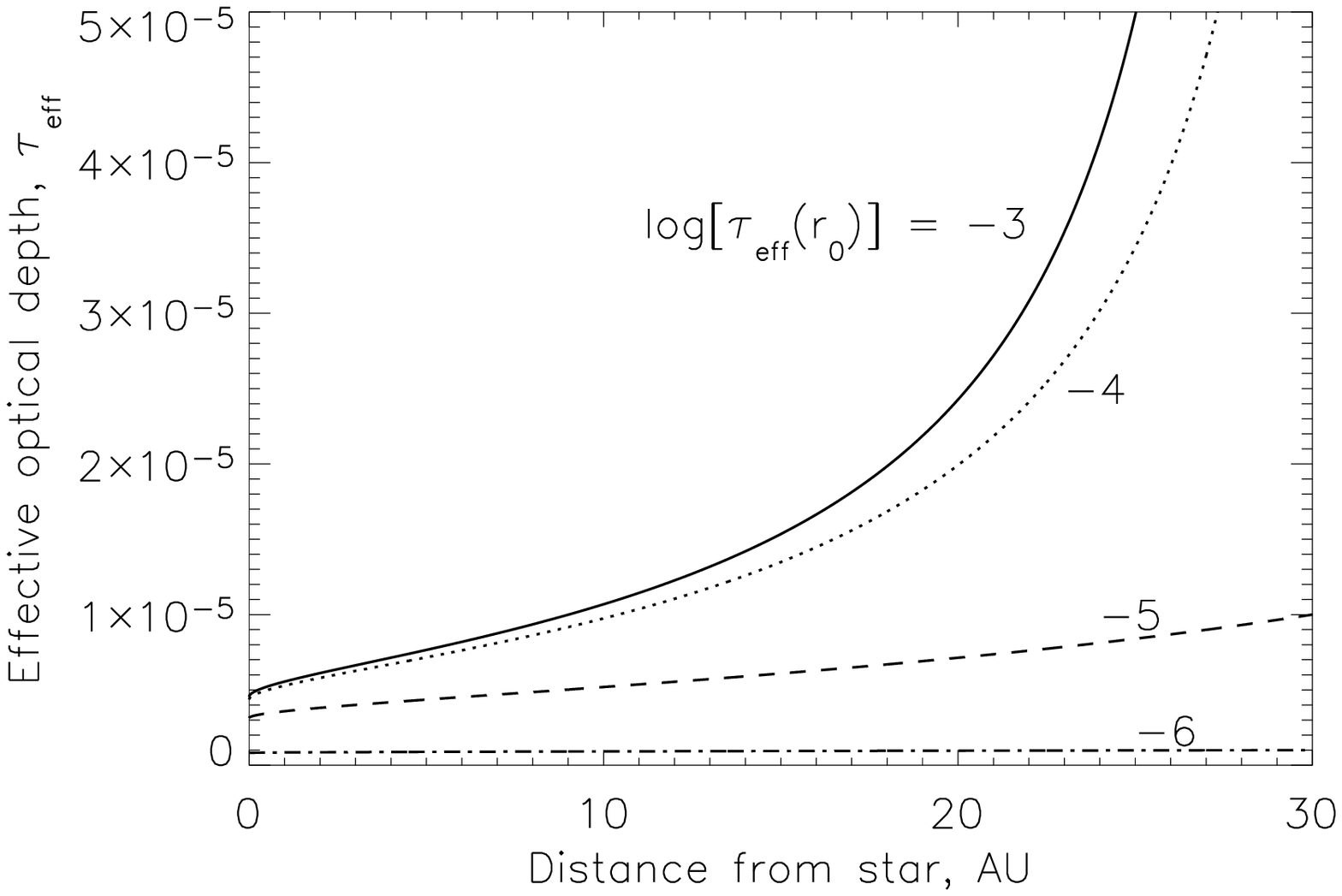}
  \end{tabular}
  \caption{The distribution of effective optical depth in a disk
  resulting from a source of same-sized particles located at $r_0$.
  The particles evolve into the inner disk due to P-R drag and mutual
  collisions (which are assumed to be destructive).
  The steady state solution depends on the parameter
  $\eta_0$ which is defined by the source parameters.
  A value of $\eta_0 = 1$ corresponds to the situation when the
  collisional lifetime of the source particles (if they suffered no
  P-R drag evolution) is equal to their P-R drag lifetime.
  \textbf{(a)} shows the functional dependence of the distribution on
  $\eta_0$.
  \textbf{(b)} shows the distribution resulting from a source of
  particles with $\beta=0.5$ located at 30 AU from a solar mass star,
  but with different dust production rates that result in
  different effective optical depths at the source.}
  \label{fig:teffeta0}
\end{figure*}

This distribution is shown in Fig.~\ref{fig:teffeta0}.
The result is that in disks which are very dense, i.e., those for
which $\eta_0 \gg 1$, most of the dust is confined to the
region where it is produced.
Very little dust in such disks makes it into the inner regions
as it is destroyed in mutual collisions before it gets there.
In disks which are tenuous, however, i.e., those for which
$\eta_0 \ll 1$, all of the dust makes it to the star without suffering
a collision.
The consequence is a dust distribution with a constant surface density
as expected from P-R drag since this is the solution to
$d[n(r)\dot{r}_{pr}] = 0$.
Dust distributions with $\eta_0 \approx 1$ have a distribution which
reflects the fact that some fraction of the dust migrates in without
encountering another dust grain, while other dust grains are destroyed.
This can be understood by considering that $\eta_0 = 1$ describes the
situation in which the collisional lifetime in the source region
given by eq.~\ref{eq:tcoll} equals the time it takes for a dust
grain to migrate from the source region to the star, which from
eq.~\ref{eq:rpr} is
$t_{pr} = 400(M_\odot/M_\star)(r_0/a_\oplus)^2/\beta$ years.

Fig.~\ref{fig:teffeta0}b shows the distribution of dust originating
in a planetesimal belt 30 AU from a solar mass star for different
dust production rates.
This illustrates the fact that the density at the centre does not
increase when the dust density reduces to a level at which P-R drag
becomes important, because even when the disk is very dense a
significant number of particles still make it into the inner system.
A look at eqs.~\ref{eq:prwcoll} and \ref{eq:eta0} shows that even in
the limit of a very dense disk the effective optical depth at the
centre of the disk cannot exceed
\begin{equation}
  max[\tau_{eff}(r=0)] = 5 \times 10^{-5} \beta
    \sqrt{(M_\star/M_\odot)(a_\oplus/r_0)},
\end{equation}
which for the belt plotted here means that the density at the centre
is at most $4.6 \times 10^{-6}$.

Of course the situation described above is a simplification, since
dust is really produced with a range of sizes.
Dust of different sizes would have different migration rates, as
defined by eq.~\ref{eq:rpr}, but would also have different collisional
lifetimes.
Eq.~\ref{eq:tcoll} was derived under the assumption that the dust
is most likely to collide with grains of similar size (Wyatt et al.
1999), collisions which were assumed to be destructive.
In reality the collisional lifetime depends on particle size, in a
way which depends on the size distribution of dust in the disk,
and the size of impactor required to destroy the particle, rather
than result in a non-destructive collision (e.g., Wyatt \& Dent 2002).
Once such a size distribution is considered, one must also consider
that dust of a given size is not only destroyed in collisions, but
also replenished by the destruction of larger particles.
The resulting continuity equation can no longer be solved
analytically, but must be solved numerically along with an appropriate
model for the outcome of collisions between dust grains of different
sizes.
Such a solution is not attempted in this paper which is more interested
in the large scale distribution of material in extrasolar planetesimal
belts for which the assumption that the observations are dominated by
grains of just one size is a fair first approximation, albeit one which
should be explored in more detail in future work.

\section{Emission Properties}
\label{s:em}
For simplicity the emission properties of the disk are derived under the
assumption that dust at a given distance from the star is heated
to black body temperatures of
$T_{bb} = 278.3 (L_\star/L_\odot)^{1/4}/\sqrt{r/a_\oplus}$ K.
It should be noted, however, that small dust grains tend to emit
at temperatures hotter than this because they emit inefficiently
at mid- to far-IR wavelengths, and temperatures above black
body have been seen in debris disks (e.g., Telesco et al. 2000).

\begin{figure*}
  \centering
     \includegraphics[width=5.0in]{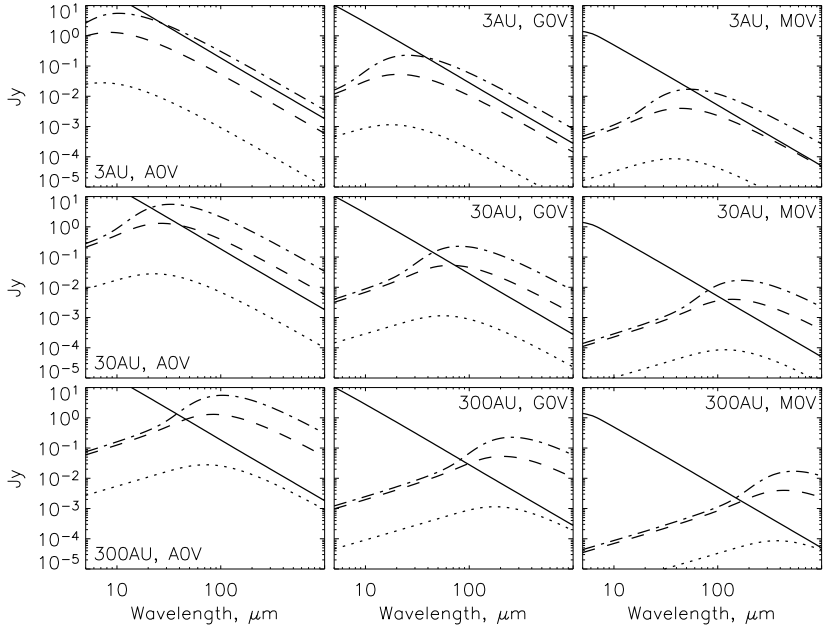} 
  \caption{The spectral energy distribution of emission from dust
  around main sequence (left) A0, (middle) G0 and (right) M0 stars.
  The top, middle and bottom rows of figures show dust from planetesimal
  belts at 3, 30 and 300 AU respectively.
  The dotted, dashed and dash-dot lines correspond to dust belts with
  $\eta_0 = 0.01$, 1 and 100 respectively, where this has been converted
  to an optical depth distribution assuming $\beta = 0.5$.
  In reality the dust may have $\beta < 0.5$, so these fluxes should
  be considered as conservative upper limits to the emission from
  these disks, even more so because black body emission efficiencies
  were assumed, whereas typical dust grains have much lower efficiencies
  at these wavelengths.
  The fluxes are normalised to stars at a distance of 10 pc, but
  naturally scale with distance squared.
  The solid line shows the stellar photosphere.}
  \label{fig:fnus}
\end{figure*}

The emission spectrum of dust from planetesimal belts around stars
of different spectral type are shown in Fig.~\ref{fig:fnus}.
The shape of these spectra can be understood qualitatively.
At the longest wavelengths all of the dust is emitting
in the Rayleigh-Jeans regime leading to a spectrum
$F_\nu \propto \lambda^{-2}$.
At shorter wavelengths there is a regime in which
$F_\nu \propto \lambda$.
This emission arises from the dust which is closest to the star.
Since dust which has a temperature $\ll 2898$ $\mu$m$/\lambda$
is emitting on the Wien side of the black body curve,
this contributes little to the flux at this wavelength.
Thus the flux at a given wavelength the comes from a region
around the star extending out to a radius $\propto \lambda^2$,
corresponding to an area $\propto \lambda^4$ and so an
emission spectrum $F_\nu \propto \lambda$ (see also Jura et al. 1998).
For dust belts in which $\eta_0 \ll 1$ the two regimes blend smoothly
into one another at a wavelength corresponding to the peak of
black body emission at the distance of $r_0$.
For more massive disks the shorter wavelength component is much smaller
leading to a spectrum which more closely resembles black body 
emission at the distance of $r_0$ plus an additional hot component.

The flux presented in Fig.~\ref{fig:fnus} includes one contentious
assumption which is the size of the dust grains used for the
parameter $\beta$.
The most appropriate number to use is that for the size of grains
contributing most to the observed flux from the disk.
In general that corresponds to the size at 
which the cross-sectional area distribution peaks.
In a collisional cascade size distribution the cross-sectional area
is concentrated in the smallest grains in the distribution.
Since dust with $\beta > 0.5$ is blown out of the system by radiation
pressure as soon as it is created, this implies that $\beta = 0.5$
is the most appropriate value to use, which is what was assumed in
Fig.~\ref{fig:fnus}.
However, evolution due to P-R drag has an effect on the size
distribution.
Since small grains are removed faster than large grains (see
eq.~\ref{eq:rpr}), the resulting cross-sectional area distribution
peaks at large sizes (Wyatt et al. 1999; Dermott et al. 2001).
Analogy with the zodiacal cloud in which the cross-sectional area
distribution peaks at a few hundred $\mu$m (Love \& Brownlee 1993)
implies that a much lower value of $\beta$ may be more appropriate,
perhaps as low as 0.01 for disks in which $\eta_0 \ll 1$.
Thus the fluxes given in Fig.~\ref{fig:fnus} should be regarded as
upper limits to the flux expected from these disks (since $\beta >
0.5$ regardless).
This is particularly true for fluxes at wavelengths longer than
100 $\mu$m, because even in a collisional cascade distribution the
emission at sub-mm wavelengths is dominated by grains larger than a few
hundred $\mu$m, since grains smaller than this emit inefficiently at
long wavelengths (see e.g., Fig. 5 of Wyatt \& Dent 2002).
Inefficient emission at long wavelengths results in a spectrum which
is steeper than $F_\nu \propto \lambda^{-2}$ in the Rayleigh-Jeans
regime.
For debris disks the observed spectrum is seen to 
fall off at a rate closer to $F_\nu \propto \lambda^{-3}$
(Dent et al. 2000).

\section{Discussion}
\label{s:disc}
A disk's detectability is determined by two factors.
First is the question of whether the disk is bright
enough to be detected in a reasonable integration time for
a given instrument.
For example, SCUBA observations at 850 $\mu$m have a limit
of a few mJy (Wyatt, Dent \& Greaves 2003)
and IRAS observations at 60 and 100 $\mu$m
had $3\sigma$ sensitivity limits of around 200 and 600 mJy.
More important at short wavelengths, and for nearby stars,
however, is how bright the disk is relative to the stellar
photosphere.
This is because unless a disk is resolved in imaging, or is
particularly bright, its flux is indistinguishable from the
stellar photosphere, the flux of which is not generally known
with better precision than $\pm 10$ \%.
For such cases an appropriate limit for detectability is that
the disk flux must be at least 0.3 times that of the photosphere.

The total flux presented in Fig.~\ref{fig:fnus} assumes that
the star is at a distance of 10 pc.
The flux from disks around stars at different distances scales
proportionally with the inverse of the distance squared.
However, the ratio of disk flux to stellar flux (shown with a
solid line on Fig.~\ref{fig:fnus}) would remain the same.
Given the constraints above, as a first approximation one can
consider that the disks which have been detected to date are
those with fluxes which lie to the upper right of the photospheric
flux in Fig.~\ref{fig:fnus}, but with the caveat that such disks
can only be detected out to a certain distance which is a function
of the instrument's sensitivity.

This allows conclusions to be reached about the balance
between collisions and P-R drag in the disks which can have
been detected.
Fundamentally this is possible because the effective optical depth
and $\eta_0$ are observable parameters (see next paragraph).
The first conclusion is that it is impossible to detect disks
with $\eta_0 \leq 0.01$ because these are too faint with respect
to the stellar photosphere.
The conclusion about disks with $\eta_0 = 1$ is less clear cut.
It would not be possible to detect such disks if they were,
like the asteroid belt, at 3 AU from the host stars.
At larger distances the disks are more readily detectable.
However, detectability is wavelength dependent, with disks
around G0V and M0V stars only becoming detectable longward
of around 100 $\mu$m, while those around A0V stars are detectable
at $>50$ $\mu$m.
Disks with $\eta_0 \gg 100$ are readily detectable for all stars,
although again there is some dependence on wavelength.

\begin{figure}
  \centering
     \includegraphics[width=3.0in]{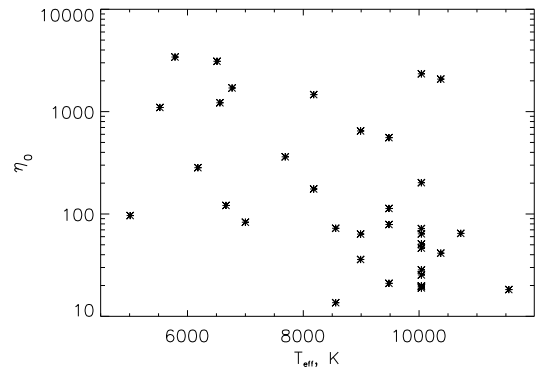} 
  \caption{The value of $\eta_0$ for the disks of the 37 stars in the
  debris disk database with excess flux measurements at two
  wavelengths plotted against the effective temperature of the
  stars.
  The disk around the star HD98800 falls off the plot at $\eta_0
  \approx 10^{5}$.}
  \label{fig:eta0teffobs}
\end{figure}

Since most disks known about to date were discovered by IRAS at
60 $\mu$m this implies that P-R drag is not a dominant factor governing
the evolution of these disks, except for perhaps the faintest disks
detected around A stars.
To check this conclusion a crude estimate for the value of $\eta_0$
was made for all disks in the debris disk database 
(http://www.roe.ac.uk/atc/research/ddd).
This database includes all main sequence stars determined in previous
surveys of the IRAS catalogues to have infrared emission in excess
of that of the stellar photosphere (e.g., Stencel \& Backman 1991;
Mannings \& Barlow 1998).
To calculate $\eta_0$, first only stars within 100 pc and with
detections of excess emission at two IRAS wavelengths were chosen.
The fluxes at the longest two of those wavelengths were then used to
determine the dust temperature and so its radius by assuming black
body emission.
Eliminating spectra which implied the emission may be associated
with background objects resulted in a list of 37 candidates, including
all the well-known debris disks.
A disk's effective optical depth was then estimated
from its flux at the longest wavelength:
\begin{equation}
  \tau_{eff} = F_\nu \Omega_{disk}/B_\nu(T), \label{eq:taueffbnu}
\end{equation}
where $\Omega_{disk}$ is the solid angle subtended by the
disk if seen face-on, which for a ring-like disk of radius $r$ and
width $dr$ at a distance of $d$ in pc is $6.8 \times 10^9 d^2/rdr$.
The ring width is generally unknown and so for uniformity it
was assumed to be $dr=0.1r$ for all disks.
Finally, $\eta_0$ was calculated under the assumption that
$\beta=0.5$.
All of these stars were found to have $\eta_0 > 10$, with a median
value of 80 (see Fig.~\ref{fig:eta0teffobs}).
\footnote{The biggest uncertainties in the derived values of $\eta_0$
are in $r$, $dr$ and $\beta$:
e.g., if black body temperatures underestimate the true radius by
a factor of 2 and the width of the ring is $dr=0.5r$ then the
$\eta_0$ values would have to be reduced by a factor of 10;
changes to $\beta$ would increase $\eta_0$.}
All 18 stars (i.e., half the sample) with $\eta_0 <80$ are of
spectral type earlier than A3V, while stars with disks with
$\eta_0 > 80$ are evenly distributed in spectral type.
It is worth noting that of the disks which have been
resolved, those with ages $\sim 10$ Myr all have $\eta_0 > 1000$
($\beta$ Pic, HR4796, HD141569) while those older than 100 Myr
all have $\eta_0 < 100$ (Vega, $\epsilon$ Eridani, Fomalhaut,
$\eta$ Corvi).

The fact that the debris disks which have been detected to date
have $\eta_0 \gg 1$ implies that the holes at their centres are
not caused by planets which prevent this dust from reaching the
inner system.
Rather the majority of this dust is ground down in mutual collisions
until it is fine enough to be removed by radiation pressure.
A similar conclusion was reached by Dominik \& Decin (2003) for   
debris disks which were detected by ISO.
This also means that azimuthal structure in the disks cannot be
caused by dust migrating into the resonances of a planet
(e.g., Kuchner \& Holman 2003), at least not due to P-R drag alone.
Models of structures in debris disks which have to invoke P-R drag
should be reconsidered and would have to include the effect of
collisions at the fundamental level to remain viable (e.g., Lecavelier des
Etangs et al. 1996), since it appears that P-R drag can effectively be
ignored in most detectable disks.

Collisions are not 100\% efficient at stopping dust from
reaching the star, and the small amount which does should result
in a small mid-IR excess.
If no such emission is detected at a level consistent with the
$\eta_0$ for a given disk, then an obstacle such as
a planet could be inferred.
However, because of the low level of this emission with respect to
the photosphere, it could only be detected in resolved imaging
making such observations difficult (e.g., Liu et al. 2004).
Even in disks with $\eta_0 \ll 1$, the resulting emission spectrum
still peaks at the temperature of dust in the planetesimal belt
itself.
This means that the temperature of the dust is a good tracer of the
distribution of the planetesimals and a relative dearth of warm
dust really indicates a hole in the planetesimal distribution
close to the star.

While uncertainties in the simple model presented here preclude
hard conclusions been drawn on whether it is possible to
detect disks with $\eta_0 \approx 1$, it is important to
remind the reader that fluxes plotted in Fig.~\ref{fig:fnus}
used the most optimistic assumptions about the amount of flux
emanating from a disk with a given $\eta_0$, so that the conclusions
may become firmer than this once a proper analysis of the evolution
of a disk with a range of particle sizes is done.
However, this study does show that detecting such disks would be much
easier at longer wavelengths, since photosphere subtraction is less
problematic here.
Disks which are too cold for IRAS to detect in the far-IR, but which
are bright enough to detect in the sub-mm have recently been
found (Wyatt, Dent \& Greaves 2003).
Thus disks with $\eta_0 \leq 1$ may turn up in sub-mm surveys of nearby
stars.
They may also be detected at 160 $\mu$m by SPITZER (Rieke et al. 2004).



\end{document}